# Experimental and Theoretical Study of Polarization-dependent Optical Transitions from InAs Quantum Dots at Telecommunication-Wavelengths (1.3-1.5μm)

Muhammad Usman[1, *, †], Susannah Heck[2], Edmund Clarke[2, Γ], Peter Spencer[2], Hoon Ryu[1], Ray Murray[2], and Gerhard Klimeck[1]

[1]Network for Computational Nanotechnology, Electrical and Computer Engineering Department, Purdue University West Lafayette IN, 47907 USA

[2]Department of Physics, Imperial College London, Blackett Laboratory, Prince Consort Road, London SW7 2AZ, UK

The design of some optical devices such as semiconductor optical amplifiers for telecommunication applications requires polarization-insensitive optical emission at the long wavelengths (1300-1550 nm). Self-assembled InAs/GaAs quantum dots (QDs) typically exhibit ground state optical emission at wavelengths shorter than 1300 nm with highly polarization-sensitive characteristics, although this can be modified by using low growth rates, the incorporation of strain-reducing capping layers or growth of closely-stacked QD layers. Exploiting the strain interactions between closely stacked QD layers also allows greater freedom in the choice of growth conditions for the upper layers, so that both a significant extension in their emission wavelength and an improved polarization response can be achieved due to modification of the QD size, strain and composition. In this paper we investigate the polarization behavior of single and stacked QD layers using room temperature sub-lasing-threshold electroluminescence and photovoltage measurements as well as atomistic modeling with the NEMO 3-D simulator. A reduction is observed in the ratio of the transverse electric (TE) to transverse magnetic (TM) optical mode response for a GaAs-capped QD stack compared to a single QD layer, but when the second layer of the two-layer stack is InGaAs-capped an increase in the TE/TM ratio is observed, in contrast to recent reports for single QD layers.

## 1. Introduction

Single and multiple InAs quantum dot (QD) layers have been explored for their potential use in the implementation of GaAs-based optical devices such as lasers, semiconductor optical amplifiers (SOAs) and saturable absorber mirrors operating at telecommunication wavelengths (1300-1550 nm) [1-4]. The optical properties of the QDs are of critical importance because they can be used to control the polarization sensitivity of the devices. Several approaches have been explored to achieve polarization-insensitive emission from QDs such as by covering the QDs with a strain reducing layer [5], growing multiple electronically-coupled layers of QDs [6-8] or vertical 'columnar' QDs [9, 10] or by formation of a type-II band alignment using GaAsSb barriers [11]. However to date there is not much theoretical guidance available to fully understand the optical properties of these QDs. Previous theoretical studies [9, 12] have explored the properties of the columnar QDs based on the k•p method which ignores atomistic granularity. The polarization properties of the QDs strongly depend on the orientation of the electron and hole wave functions which are determined by the asymmetric nature of the interface between the QD and the surrounding GaAs buffer, strain and piezoelectric fields [13-15]. Any quantitative analysis of the quantum dot devices to fully incorporate all of the mentioned effects must involve modeling and simulations of such nanostructures at atomistic scale. The studies based on the continuum methods such as the effective mass model or k•p [10, 12] that lack the atomistic resolution cannot include the interface roughness, alloy randomness, and strain-induced symmetry lowering.

This work explores the wavelength and polarization properties of independent layers (hereafter referred to as single quantum dot layers (SQD)) and two closely-stacked layers (bilayers) of InAs/GaAs QDs incorporated into ridge-waveguide laser structures, using room temperature (RT) sub-lasing-threshold electroluminescence (EL) and photovoltage (PV) measurements and associated atomistic modeling of the whole sample geometry. The experimental data is analyzed and explained by atomistically modeling the QD geometries using the NEMO 3-D simulator [16, 17]. Our results indicate that telecommunication wavelengths can be achieved by growing QDs in the form of vertical stacks without incorporating dilute nitride layers or forming the columnar QDs. The polarization-resolved measurements and theoretical calculations show a reduction in the ratio of the transverse electric (TE) to transverse magnetic (TM) optical mode response for a GaAs-capped InAs QD bilayer compared to a single GaAs-capped InAs QD layer.

In contrast to a recent study on the single InAs QD layers [5], our experimental and theoretical results indicate an increase in the TE/TM ratio when the upper QD layer in the bilayer is covered by a InGaAs strain-reducing capping layer (SRCL). This increase in the TE/TM ratio is due to the biaxial strain induced HH-LH splitting, which increases in the presence of the SRCL. This result is is consistent with our earlier study of single InAs QD layers capped by an InGaAs SRCL [13]. Our experimental measurements and theoretical calculations indicate that the InGaAs SRCL can only red shift the optical wavelength and does not reduce the TE/TM ratio for isotropic polarization response.

## 2. Experimental Setup

The devices investigated are ridge-waveguide QD laser structures with 500 nm thick GaAs active regions incorporating a series of three or five QD bilayers, and a reference sample with an active region containing five single QD layers, separated by 50 nm. The active region was surrounded by 1500 nm $Al_{0.3}Ga_{0.7}As$ $n$- and $p$-doped cladding layers. For the sample containing single layers, each QD layer was grown by annealing of the GaAs surface under an As overpressure at 580 °C for 10 minutes to minimize surface undulations then deposition of 2.4 ML InAs at a temperature of 485 °C and at a growth rate of 0.014 $MLs^{-1}$. The QDs were capped by 15 nm GaAs grown at 492 °C before the remaining 35 nm GaAs cap was grown at 580 °C. These growth conditions yield QD layers with a QD density of 1.5 x $10^{10}$ $cm^{-2}$. For the GaAs-capped bilayers, the first (seed) layer of QDs was formed by deposition of 2.4 ML InAs at a temperature of 480 °C and a growth rate of 0.014 $MLs^{-1}$ (similar conditions to those for the single QD layers). These QDs were then capped by a 10 nm GaAs spacer layer, also grown at 480 °C and the sample was then annealed under an As overpressure at 580 °C or 10 minutes, to reduce surface undulations and also to desorb segregated In from the underlying QD layer [18]. The second QD layer was then formed by deposition of 3.3 ML InAs at a lower growth temperature of 467 °C. These QDs were then capped by 15 nm GaAs also at 467 °C before the remaining 35 nm GaAs was grown at 580 °C. The reduced growth temperature for the second layer of the bilayer is crucial for achieving the extended emission wavelength and high uniformity of the QDs by suppression of strain-induced intermixing effects [19, 20]. The QD density in each layer of the bilayer is 2.7 x $10^{10}$ $cm^{-2}$, similar to the single QD layers. To achieve the maximum extension of the emission wavelength, growth conditions for the InGaAs-capped bilayers were modified compared to the GaAs-capped bilayers, with the seed layer now grown at 505 °C, leading to a lower density (5 x $10^9$ $cm^{-2}$) of larger QDs. The reduced density and increased size of the QDs in the seed layer leads to a concomitant increase in the size of the QDs in the second layer [21]. Also, the second QD layer was now capped by 4 nm $In_{0.26}Ga_{0.74}As$ then 11 nm GaAs at 467 °C before growth of the remaining 35 nm GaAs at 580 °C. These growth conditions lead to an extension of the emission wavelength to 1470 nm at room temperature; this is a shorter wavelength than previously reported for individual InGaAs-capped QD bilayers [21] but the growth conditions for multiple closely-stacked QD bilayers have not yet been optimized. PV measurements were obtained by illuminating the front facet of the devices with either TE or TM-polarized light from a lamp dispersed by a monochromator [22].

## 3. Theoretical Model

The cross-sectional transmission electron microscopy (TEM) images in figure 1 (a) and (b) indicated that the lower quantum dot in the bilayer sample has a diameter of ~20nm and height of ~7nm. The upper quantum dot with the GaAs cap (figure 1(a)) has a diameter of ~30nm and a height of ~8nm. The InGaAs cap tends to increase the height of the upper quantum dot due to reduced out-diffusion of indium from the QDs during capping [20, 21, 23]. This results in slightly taller QDs with a height of ~10nm as can be seen in figure 1 (b).

Schematics of the model QDs in the SQD and bilayers are shown in figure 1 (c, d, e). We consider in figure 1(c) an InAs single QD embedded inside a GaAs buffer, in figure 1(d) a bilayer consisting of two InAs QDs embedded inside a GaAs matrix, and in figure 1(e) a bilayer consisting of two InAs QDs embedded inside a GaAs matrix but with the upper QD capped with a 4nmm $In_{0.26}Ga_{0.74}As$ SRCL. The dimensions of the SQD are same as those of the quantum dots in the lower layer of the bilayers.

The theoretical calculations are performed using the NEMO 3-D simulator [16, 17] which calculates the electronic structures through multi-million atom simulations based on a twenty band $sp^3d^5s^*$ nearest neighbor empirical tight binding model [24]. The strain is calculated using an atomistic Valence Force Field (VFF) model, with the Keating potential modified to include the anharmonic corrections [25]. Both linear and quadratic piezoelectric potentials [14, 26] are included. The inter-band optical transition strengths are calculated using Fermi's golden rule as the squared magnitude of the momentum matrix elements summed over the spin degenerate states [26]. We have used large strain domains to fully incorporate the long range impacts of strain and piezoelectric fields. For the bilayers, the strain buffer has dimensions 70 nm x 70 nm x 66 nm, containing approximately 20 million atoms. The strain domain has fixed and free boundary conditions on the bottom and top respectively, whereas periodic boundary conditions are used in the lateral directions. The electronic domain is relatively small, extending 60 nm laterally and having a height of 50 nm in the [001] direction. We implemented closed boundary conditions for the electronic domain in all of the three dimensions passivated by our published model [27].

We emphasize here that all the simulations are performed using previously published VFF model constants [28], tight binding parameters [24], and piezoelectric constants [29]. The only "inputs" to the device simulator are the geometries and the material compositions as indicated in figure 1 (c, d, e). The close agreement between calculations and experimental results obtained, without any adjustments in the published parameters, documents the transferability of the empirical VFF and tight binding parameters similar to previous work on InAs/GaAs quantum dots [13, 26] and SiGe [30].

## 4. Results and Discussions

*Only the upper QD is optically active:* In bilayer QDs, the upper QD is slightly larger than the lower QD [31]. Strain due to the upper QD tends to push the lower QD energy levels to the higher values and thus the lowest electron and highest hole energy levels reside in the upper QD [32]. Due to a large separation between the quantum dot layers (10nm), the electron and hole wave functions are confined in the upper quantum dot and do not form hybridized molecular states. Such hybridized states can be observed for closely stacked QDs, separated by ~6nm or less [26, 32], or can be observed by applying external electric fields [26, 33]. In both bilayers under study the first three electron and hole energy levels are confined to the upper QD. The first electron energy level in the lower QD is the fourth energy level, E4, of the structure in the case of the bilayer with the GaAs-cap, and the sixth energy level, E6, in the case of the bilayer with the InGaAs-SRCL. In these bilayers the upper QD serves as an optically active layer for the ground state optical emission. The lower QD does not contribute to optical emission for reasonably low carrier occupation in agreement with previous photoluminescence (PL) measurements [20, 21, 34].

*Red shift of optical emissions to 1300 nm and beyond:*

Figure 2 compares the RT EL spectra obtained from the SQD (red curve) to that measured on a bilayer without a SRCL (black curve). A low current injection level (current densities of less than 25 $Acm^{-2}$) was used in order to suppress heating effects and amplified spontaneous emission. The dotted lines are the electron-hole transition energies computed from the NEMO 3-D simulations, which closely match the experimental data. A red shift of ~75 nm is observed for the bilayer compared to the SQD sample. This is because of the larger height of the upper QD and the strain coupling between the two layers of the QD stack which tend to reduce the optical gap and red shifts the optical emission [32]. Figure 2 also compares the RT EL spectra measured on the bilayers with (green) and without (black) the InGaAs-SRCL. The InGaAs-SRCL relaxes the hydrostatic strain and reinforces the biaxial strain which causes a reduction in the optical gap and further red shift in the emission wavelength [13]. A ~122 nm red shift is observed induced by the InGaAs-SRCL.

*Physics of the InGaAs-SRCL induced red shifts:* Figure 3 compares the local band edges of the lowest conduction band (CB) and two highest valence bands (HH and LH) for the two bilayers under study (figures 1(b) and (c)). The InGaAs-SRCL reduces the hydrostatic strain and increases the biaxial strain in the upper QD layer thus shifting the CB band edge to the lower energies and the HH band edge to the higher energies [13]. This results in a reduction of ~50 meV in the band gap inducing ~88 meV reduction in the optical gap. As a result, the ground state optical emission red shifts by ~122 nm. Thus ground state optical emission above 1300 nm can be achieved from the bilayer QDs.

*Hole energy levels in the (110) and (-110) directions:* Figure 4 shows a top view of the spatial distribution of the lowest conduction band energy level (E1) and three highest hole energy levels (H1, H2, and H3) for the SQD sample (top row) and the bilayer QD without a InGaAs-SRCL (bottom row). The electron energy state E1 possesses the symmetric distribution of an s-type wave function. However, the hole states are oriented along the [110] or [-110] directions due to strain and piezoelectric field induced symmetry lowering [13-15, 35, 36]. Figure 4 shows the plot of wave functions for only one state corresponding to the Kramer's doublet. The other degenerate state will have the wave function concentrated on the opposite edge of the QD. As mentioned earlier in the methodology section, the optical transition rates are calculated by adding contribution from the both degenerate states. It should also be noted that all of the first three hole wave functions plotted are concentrated at the interface of the QD rather than being at the QD center. This is due to the large aspect ratios (height/base) of these quantum dots which results in an increase in biaxial strain at the QD/GaAs matrix interface, creating HH traps (pockets) [35, 36].

Due to the anisotropy of the hole wave functions along the [110] and [-110] directions, the inter-band optical transitions between the electron and hole states, namely E1↔H1, E1↔H2, and E1↔H3, will be strongly polarization dependent. However, polarization-resolved photoluminescence collected from surface of equivalent unprocessed samples indicate the emission is isotropic in the plane of the QDs *i.e.* TE(110)/TE(-110) ~ 1.0 +/- 0.1. Similar results were reported earlier [37], where an isotropic polarization response was observed in the plane of the quantum dot for slightly smaller QDs having similar aspect ratio ~0.3. The reason for the discrepancy between theory and experiment was unclear and the authors [37] reported that it may be due to "shape variations or possible omissions in the theories" resulting in an anisotropic response.

We demonstrate here that in order to achieve the measured isotropic polarization dependence in the plane of the QD, more than one hole energy levels should be included in the calculation of the ground state optical transition strength. This is because the hole energy levels are very closely packed, as can be seen in figure 2 where the transition energies between the ground electron state E1 and the first three hole energy levels (H1, H2, and H3) are plotted with the experimentally measured optical spectra.. The difference between the highest and the lowest transition energies calculated here (E3-H1)-(E1-H1) is ~16 meV, ~12.5 meV, and ~14 meV for the SQD, the bilayer with the GaAs cap, and the bilayer with the InGaAs-SRCL, respectively. This shows that the first three hole energy levels in such QDs are approximately within $0.5k_BT$ (~ 12.9 meV) at 300 K. It can be concluded that in such QD systems, the top most valence band states are very closely packed within few meV of the energy range. This implies that at RT, multiple hole levels can contribute to the measured transition intensity. In the next section, we will show that multiple hole energy levels can indeed simultaneously contribute to the ground state optical emission and hence must be considered in any theoretical calculation of the ground state optical transition strengths.

*Multiple holes are required to achieve an isotropic in-plane polarization:* Figure 5 plots the optical intensity computed from the NEMO 3-D simulations as a polar plot for (a) the SQD and (b) the bilayer without the InGaAs-SRCL. The direction of the incident light is along the [001] direction. The optical transition intensities are plotted as a function of the angle 'θ' between the [100] and the [010] directions. For the SQD case (figure 5(a)), the simulations give a slightly anisotropic polarization dependence i.e. TE(110)/TE(-110) ~ 1.18. This is because the first three hole energy levels are oriented along [110] (see top row of figure 4), which results in the TE(-110) mode being slightly weaker than the TE(110) mode. Figure 5(b) shows nearly isotropic polarization emission for the GaAs-capped bilayer. This is because H1 is oriented along [110] direction, while H2 and H3 are both oriented along [-110] direction. This orthogonal distribution of the hole wave functions tends to cancel out the polarization sensitivity and hence the resultant optical spectrum become nearly polarization insensitive in the plane of the QD. The simulations compute TE(110)/TE(-110) ~ 1.07. We conclude that any theoretical study of the polarization-resolved ground state optical emission must include multiple hole energy levels to achieve the polarization insensitivity in the plane of the QD. Past studies [12, 38] only consider the topmost valence band state.

*TE/TM ratio analysis:* For telecommunications applications, a polarization insensitive response is desirable for some edge-emitting devices. QDs grown by the self-assembly process have low aspect ratios, typically in the range of 0.1 to 0.3, and possess a strong confinement of the charge carriers in the [001] direction. This results in very anisotropic optical spectra where the TE optical mode is dominant and TM optical mode is very weak. The QDs having aspect ratios larger than 0.6 are designed for isotropic polarization response in the form of large QD stacks with closely packed QDs known as the 'columnar QDs' [9, 10]. Here we analyze the TE and TM polarized emission for a SQD and compare it with the bilayers with and without an InGaAs-SRCL as shown in the figure 1 (c, d, e), using cleaved-edge PV measurements and calculation of the polarization dependent optical transitions.

*Single QD to Double QD without SRCL → TE/TM ratio decreases:* Figure 6 compares the optical intensity model results computed from the NEMO 3-D simulator and represented as polar plots for the SQD (red solid curve), the bilayer without the InGaAs-SRCL (black dotted curve), and the bilayer with the InGaAs-SRCL (green solid line). The direction of the incident light is along the [-110] direction. The inter-band optical transition intensities are calculated as a function of the angle θ between [001] and [110]. Each curve represents the sum of the optical intensities of the E1↔H1, E1↔H2, and E1↔H3 transitions. Figure 6(a) shows that the TE/TM ratio decreases for the bilayer when compared to the SQD device. This is because in a bilayer system, the strain of the lower QD takes part in the growth of the upper QD and results in an increase in the size of the upper QD. The taller upper QD reduces the [001] carrier confinement and hence the TM optical mode is enhanced, leading to a reduction in the TE(110)/TM(001) ratio.

*Double QD to Double QD with SRCL → TE/TM ratio increases:* Figure 6(b) compares the polar plots of the bilayer without and with the InGaAs-SRCL. A significant increase in the polarization sensitivity occurs when the upper QD layer is covered by the InGaAs-SRCL. The reason for this increase in the TE(110)/TM(001) ratio can be understood with the help of the valence band edge diagrams previously shown in figure 3. This figure shows that the InGaAs-SRCL shifts the HH and LH band edges in the opposite directions due to the biaxial strain reinforcement [13] leading to an increase in the splitting between the two valence bands. As a result, the top most valence band (hole) states will have enhanced HH character. We calculate that the HH/LH ratio increases from ~24.41, ~23.23, and ~21.38 to ~25.43, ~23.30, and ~22.19 for H1, H2, and H3, respectively. This increases the TE(110) and suppresses the TM(001) optical mode resulting in an increase in the TE(110)/TM(001) ratio.

These results for bilayer with SRCL are in contrast to a recent experimental study [5] where the InGaAs-SRCL is shown to decrease the TE/TM ratio for a single InAs QD. That single quantum dot result was attributed to the significant increase in the QD aspect ratio, from 0.235 to 0.65, due to the influence of the InGaAs cap in preserving the QD height during the capping process. For the bilayer QDs, the relatively low capping temperature for QDs in the second layer will also preserve the QD height [20] and we do not observe a significant further enhancement in the QD aspect ratio due to InGaAs capping in this case. The TEM images shown in the figure 1 (a) and (b) clearly indicate that InGaAs SRCL does not drastically change the aspect ratio of these quantum dots, in contrast to the single-layer InGaAs-capped QDs reported in Jayavel *et al.* [5].

The table shown in the figure 6(c) summarizes the comparison of the TE/TM ratios from the experimental PV measurements and the NEMO 3-D based calculations for the three QD systems. The calculated values are in reasonable agreement with the experimental measurements for all of the three cases and show similar trends.

## 5. Conclusions

We have experimentally and theoretically investigated the RT optical properties of single InAs/GaAs QD layers and InAs/GaAs QD bilayers. Ground state optical emission at wavelengths in excess of 1300 nm is achieved from the QD bilayers. The PV measurements and optical transition strength calculations indicate reduced polarization sensitivity for the bilayer QD stack covered with a GaAs-cap when compared to the single QD layer. However, the InGaAs-SRCL increases the polarization dependence due to biaxial strain enhancement. This is in contrast to a recent experimental study [5] where InGaAs-SRCL is shown to reduce the polarization sensitivity for independent QD layers. The comparison of the optical model calculation with the surface PL measurements emphasizes that more than one hole energy levels must be included in the calculations of the ground state optical spectra to achieve isotropic polarization sensitivity in the plane of the QD. Atomistic strain, piezoelectricity, electronic structure, and optical transition strength calculations are in reasonable agreement with the experimental measurements and help to understand the optical properties of such QDs systems.

## *ACKNOWLEDGEMENTS*

The TEM images in the figure 1 (a, b) are courtesy of Dr Richard Beanland, Integrity Scientific (www.integrityscientific.com) and Prof. Richard Hogg, University of Sheffield. This work uses the computational resources provided by the National Science Foundation (NSF) funded Network for Computational Nanotechnology (NCN) through http://nanohub.org, and RCAC at Purdue University. Hoon Ryu and Gerhard Klimeck acknowledge NSF funds of grant EEC-0228390 and ECCS-0701612. Muhammad Usman acknowledges the funding from the Fulbright fellowship sponsored by the USA Department of States under grant 15054783. Muhammad Usman also acknowledges useful discussions with Prof. Eoin O'Reilly (Tyndall National Institute) and Prof. Timothy B. Boykin (University of Alabama) about the calculation of HH and LH contributions in the valence band states.

# REFERENCES:

*Corresponding Author: usman@alumni.purdue.edu  
†Now at Tyndall National Institute, Lee Maltings, Cork, Ireland.  
ˡNow at EPSRC National Centre for III-V Technologies, Centre for Nanoscience and Technology, University of Sheffield, North Campus, Broad Lane, Sheffield S3 7HQ, U. K.

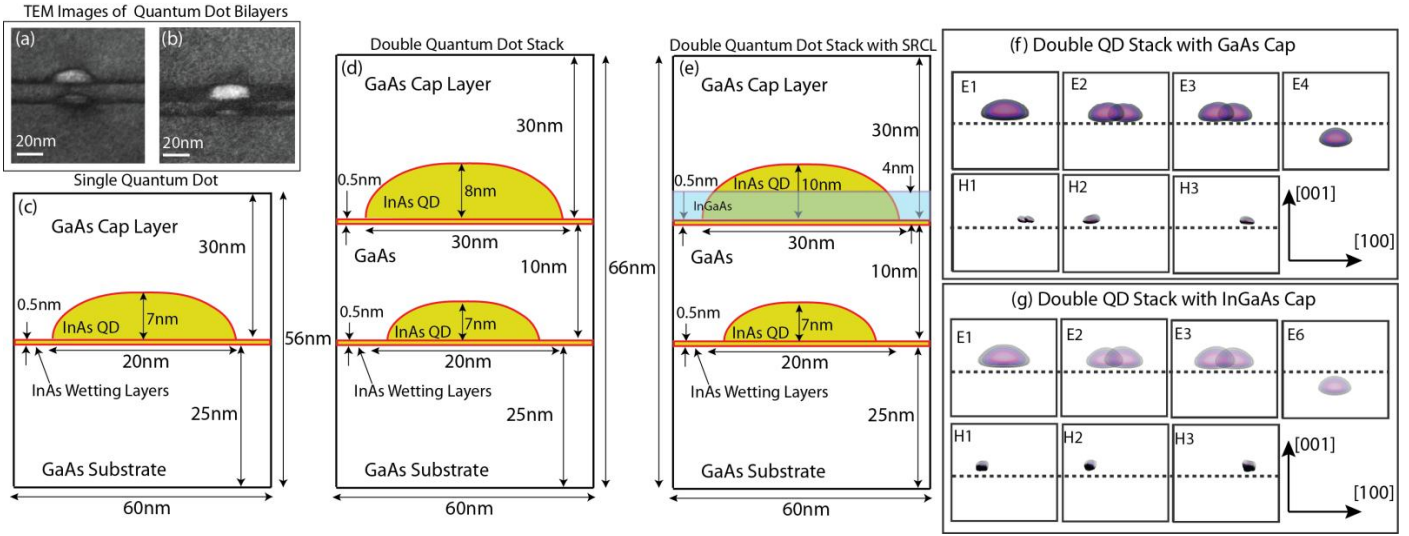

Figure1: (a, b) TEM Images of the bilayer QD stacks without (a) and with (b) the SRCL capping. The SRCL overgrowth clearly tends to increase the size of the QD. (c) Geometry of a single InAs QD lying on top of a 0.5 nm InAs wetting layer inside GaAs buffer. (d) Geometry of a two InAs QD vertical stack surrounded by GaAs buffer. Both QDs are lying on the top of 0.5 nm wetting layers. The separation between the wetting layers is 10 nm. The upper QD in the stack is larger than the lower QD due to strain driven self-assembly process. (e) Geometry of a two InAs QD vertical stack surrounded by GaAs buffer. Both QDs are lying on the top of 0.5 nm wetting layers. The separation between the wetting layers is 10 nm. The upper QD is first covered by a 4nm $In_{0.26}Ga_{0.74}As$ strain reducing layer before depositing the GaAs capping layer. The upper QD has slightly larger size than the upper QD of the stack without SRCL (part (d)) due to reduced In segregation effect. (f) Wave function plots of first four electron and first three hole energy levels of the two QD stack (part (e)). The dotted line is marked to guide the eyes and separates the upper and the lower QDs. All of the first three electrons and first three holes are in the upper QD indicating that the upper QD is optically active where as the lower QD remain inactive. (g) Wave function plots of four electron and three hole energy levels of the two QD stack with the SRCL cap (part (e)). The dotted line is marked to guide the eyes and separates the upper and the lower QDs. All of the first three electrons and first three holes are in the upper QD indicating that the upper QD is optically active whereas the lower QD remains inactive and serves to control the optical properties of the upper QD through the strain coupling. The first electron state in the lower QD is $E_6$.

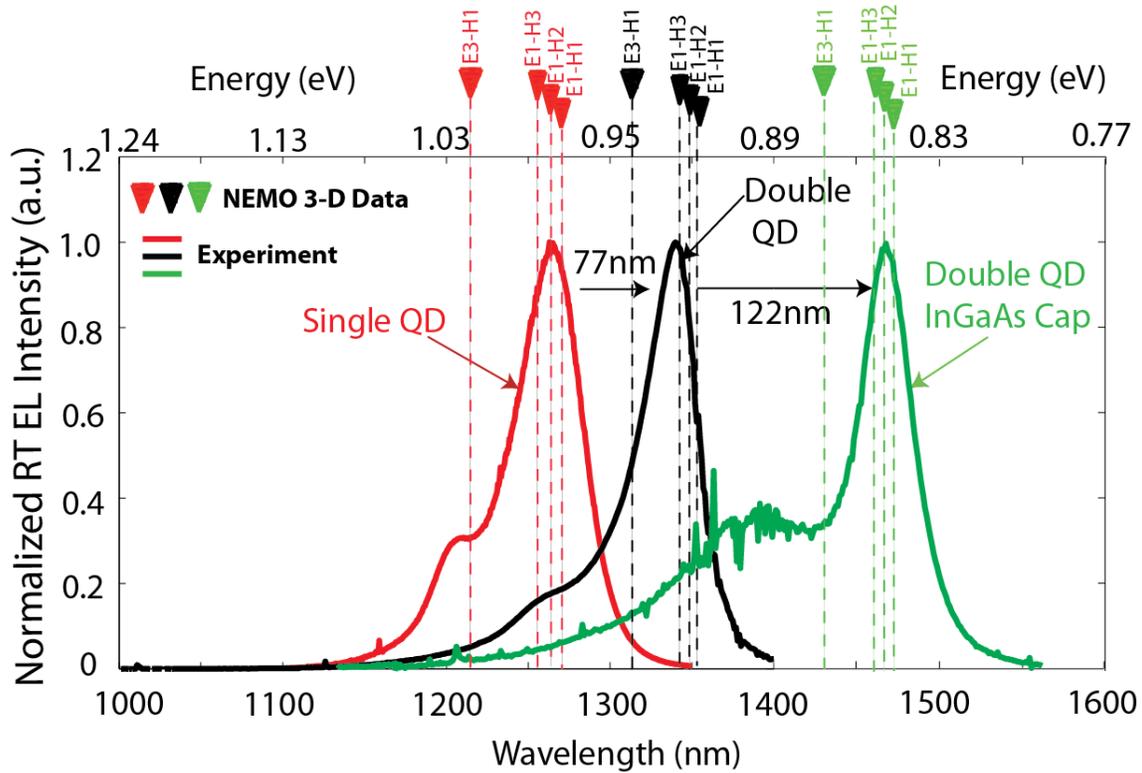

Figure 2: Comparison of the ground state optical emissions from the single InAs QD and the double InAs QD stack without SRCL and with SRCL. The red (single QD), black (QD stack without the SRCL) and green (QD stack with the SRCL) curves show the room temperature electroluminescence spectra measured in the experiment. The vertical dotted lines are the electron-hole transition energies calculated from the NEMO 3-D simulations. A red shift of ~77 nm is observed for the bilayers without the SRCL compared to the single InAs QDs. The InGaAs SRCL further red shifts the spectrum by ~122 nm. The noise in the green color close to 1400 nm wavelength is due to water absorption.

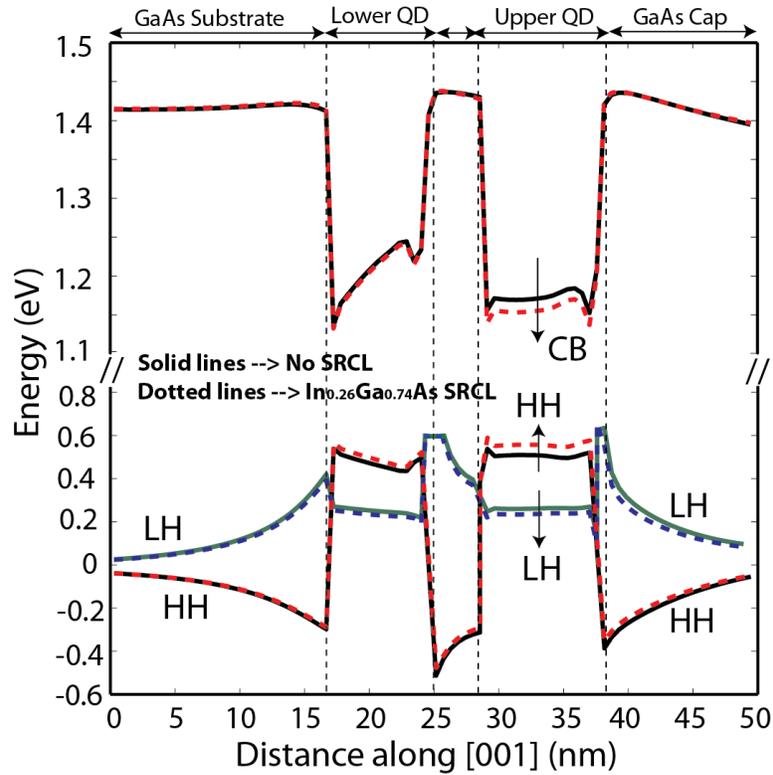

Figure 3: The comparison of the lowest conduction band edge (CB) and the highest two valence band edges (HH and LH) for the bilayer QD stacks with (dotted lines) and without (solid lines) the SRCL. The InGaAs SRCL shifts the CB edge to lower energies and the HH band edge to the higher energies, thus reducing the band gap and increasing the optical emission wavelength. The SRCL also shifts the HH and LH band edges in the opposite directions and increases the HH-LH separation. This results in an increase in the TE/TM ratio.

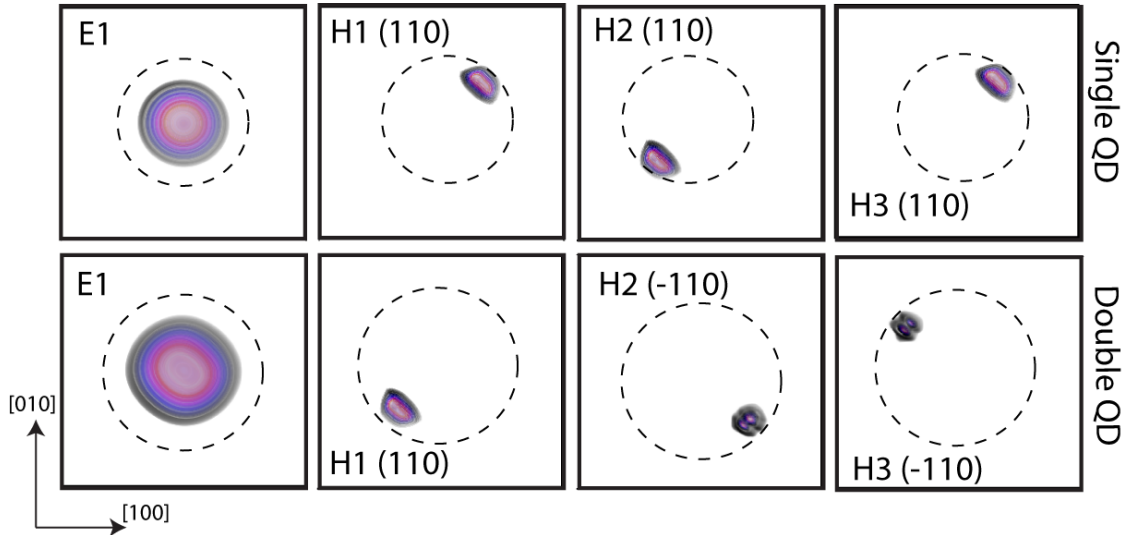

Figure 4: Top view of the spatial distribution of the lowest electron energy level (E1) and the highest three hole energy levels (H1, H2, and H3) for the single QD (upper row) and the double QD stack without the SRCL (lower row). Dashed circles are drawn to highlight the boundaries of the QD regions. The lowest electron energy level is of s-type character and shows a symmetric distribution of the charge density. The hole energy levels are strongly affected by the strain and piezoelectricity and tend to align along the [110] or [-110] directions.

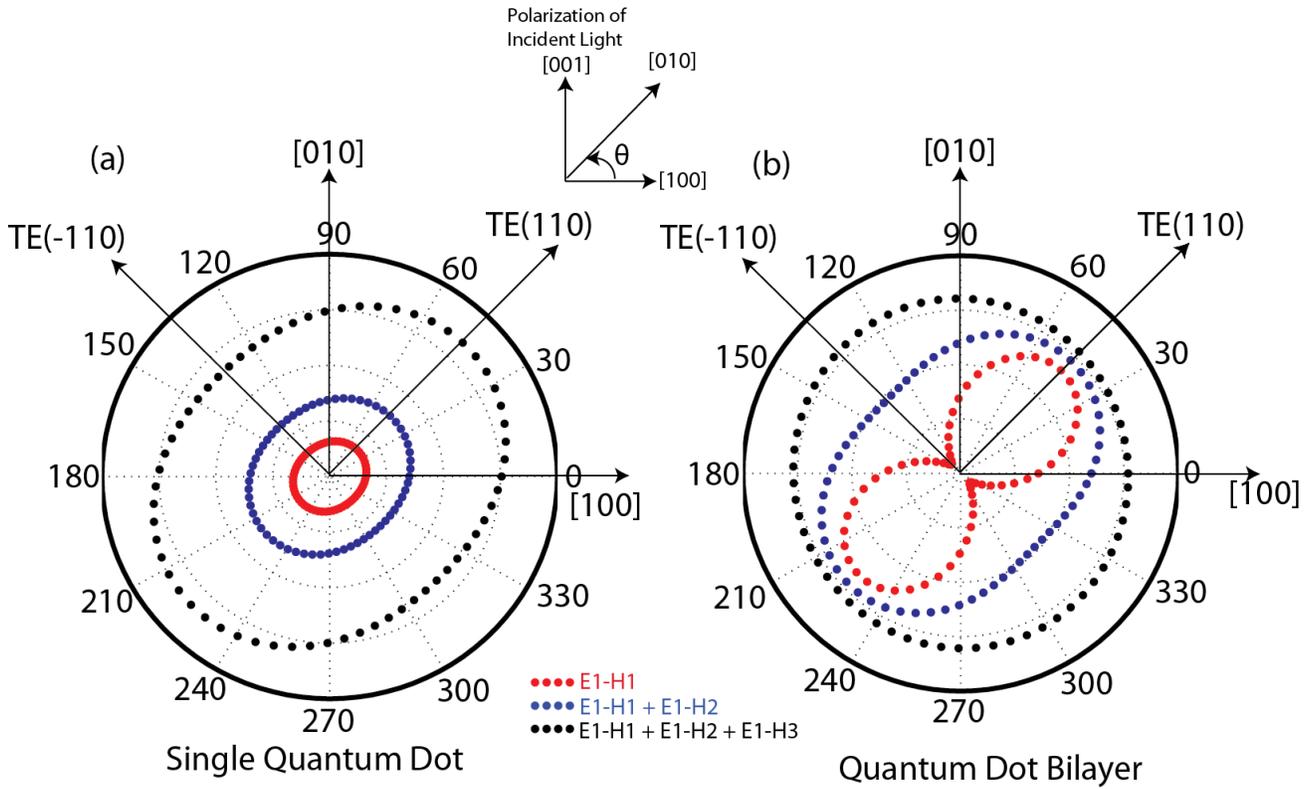

Figure 5: Optical intensity model results represented as polar plots for (a) the single InAs QD and (b) the bilayer InAs QD stack without the SRCL. The direction of the polarization of the incident light is assumed along the [001] direction. The optical transition intensities are plotted as a function of the angle θ between the [100] and the [010] directions. Red dots: only E1↔H1 transition is plotted. Blue dots: the sum of E1↔H1 and E1↔H2 transitions is plotted. Black dots: the sum of E1↔H1, E1↔H2 and E1↔H3 transitions is plotted.

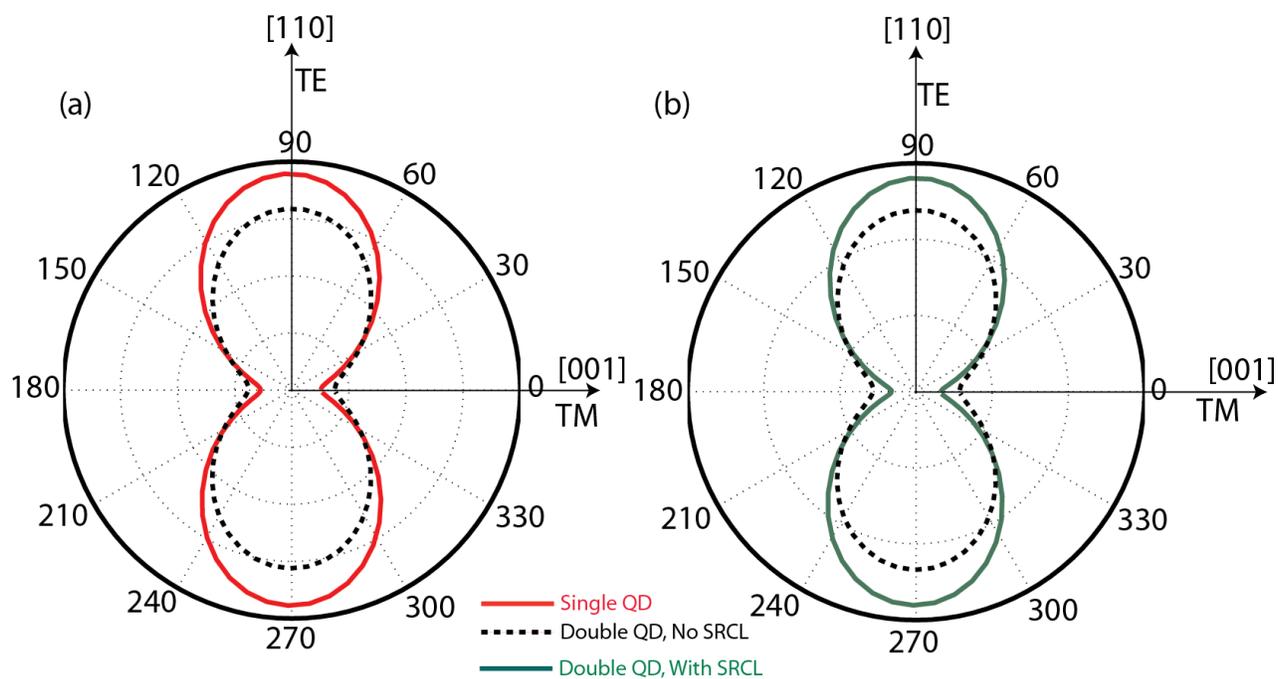

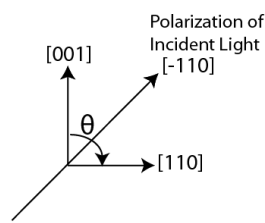

Figure 6: (a, b) Optical intensity model results represented as polar plots are shown for the single InAs QD (red solid curve), the bilayer InAs QD stack without SRCL (black dotted curve), and the bilayer InAs QD stack with SRCL (green solid line). The direction of the polarization of the incident light is assumed to be along the [-110] direction. The inter-band optical transition intensities are calculated as a function of the angle θ between the [110] and the [001] directions. Each curve represents the sum of the intensities of the E1↔H1, E1↔H2, and E1↔H3 transitions. (c) The comparison of the TE(110)/TM(001) ratios for the three QD systems from experimental PV measurements and NEMO 3-D calculations.